\documentclass[manuscript]{aastex}
\usepackage{epstopdf}

\newcommand{\mps}{m~s$^{-1}$}
\newcommand{\kmps}{km~s$^{-1}$}

\shorttitle{}
\shortauthors{Crockett et al.}

\begin{document}

\title{Precision radial velocities with CSHELL}
\author{Christopher J. Crockett \altaffilmark{1,2,3}, \\
	Naved I. Mahmud \altaffilmark{3,4}, \\
	L. Prato \altaffilmark{1,3}, \\
	Christopher M. Johns-Krull \altaffilmark{3,4}, \\	
	Daniel T. Jaffe \altaffilmark{5}, \\
	Charles A. Beichman \altaffilmark{6,7}}

\altaffiltext{1}{Lowell Observatory, 1400 W Mars Hill Road, Flagstaff, AZ 86001; \\	
	crockett@lowell.edu, lprato@lowell.edu}
\altaffiltext{2}{Department of Physics and Astronomy, University of California Los Angeles, \\
	430 Portola Plaza, Box 951547, Los Angeles, CA 90095-1547}
\altaffiltext{3}{Visiting Astronomer at the Infrared Telescope Facility, which is operated by the University of Hawaii under Cooperative Agreement no. NCC 5-538 with the National Aeronautics and Space Administration, Science Mission Directorate, Planetary Astronomy Program.}
\altaffiltext{4}{Department of Physics and Astronomy, Rice University, MS-108, 6100 Main Street, \\
	Houston, TX 77005; naved@rice.edu, cmj@rice.edu}
\altaffiltext{5}{Department of Astronomy, University of Texas, R.L. Moore Hall, Austin, TX 78712; dtj@astro.as.utexas.edu}
\altaffiltext{6}{Jet Propulsion Laboratory, California Institute of Technology, 4800 Oak Grove Drive, Pasadena, CA 91109}
\altaffiltext{7}{NASA Exoplanet Science Institute (NExScI), California Institute of Technology, 770 S. Wilson Ave, Pasadena, CA 91125}

\begin{abstract}

Radial velocity identification of extrasolar planets has historically been dominated by optical surveys.  Interest in expanding exoplanet searches to M dwarfs and young stars, however, has motivated a push to improve the precision of near infrared radial velocity techniques.  We present our methodology for achieving 58 \mps~precision in the K band on the M0 dwarf GJ 281 using the CSHELL spectrograph at the 3-meter NASA IRTF.  We also demonstrate our ability to recover the known 4 $M_{JUP}$ exoplanet Gl 86 b and discuss the implications for success in detecting planets around 1 $-$ 3 Myr old T Tauri stars.
\end{abstract}


\section{Introduction}
\label{sec:Intro}
Since 1989, astronomers have discovered over 500 extrasolar planets, $\sim$92\% of which were identified using radial velocity (RV) techniques\footnote{Data from \emph{The Extrasolar Planet Encyclopedia} (www.exoplanet.eu)}.  Increased time baselines and improved velocity precision \citep[$\sim$1 \mps~e.g.,][]{Mayor:03:20} have led to an abundance of ground-based RV planet discoveries including the lightest known planet around a main sequence star \citep[$m \sin i$ = 1.7 $M_{\earth}$,][]{Mayor:09:487} and the recent claim of a $\sim$3 $M_{\earth}$ planet orbiting within its star's habitable zone \citep{Vogt:10}.  Ongoing exoplanet surveys typically focus on slow-rotating FGK dwarfs.  The numerous, narrow atomic lines in the photospheres of these stars allow for precise RV determination.  The SEDs of these stars also peak in the optical region of the spectrum, facilitating their study with optical detectors, a more mature technology than is available in other wavelength regions, such as near-infrared (NIR). 

The past few years, however, have seen a surge of interest in applying precision RV techniques in the NIR, in part to explore M dwarfs as potential planet hosts.  These late-type stars are faint at optical wavelengths but comparatively bright at 1 $-$ 2 \micron.  They are also the most numerous stars in our Galaxy; by focusing on FGK dwarfs in most surveys to date the majority of stars have been neglected.  Also, for a given orbital period and planet mass, the RV amplitude scales with host mass as $M_*^{-2/3}$.  Therefore, planets should be easier to detect around lower mass M and L dwarfs.  Furthermore, including the low mass stars expands the available parameter space to include lower mass planets.  The M dwarfs are also interesting targets to the astrobiology community \citep{Tarter:07:30}.  Because of their lower luminosity, the habitable zones (HZ) are significantly closer to the central star than for higher mass stars and the RV amplitudes of HZ planets are correspondingly greater, thereby increasing the possibility of discovering a habitable Earth-mass planet \citep{Vogt:10}.  

M dwarfs may also serve as testing grounds for models of planet formation.  In spite of all the planet discoveries in the past twenty years, a unified model of planet formation is still elusive.  In the core accretion model \citep[e.g.,][]{Pollack:96:62}, a planetary core is built up through the accretion of ices, dust particles, and eventually planetesimals in the circumstellar disk.  \citet{Laughlin:04:L73} present calculations which indicate that giant planet formation around M dwarfs under a core-accretion paradigm is inhibited as a result of lower surface densities in the disk beyond the snowline, where Jupiter-mass worlds are most likely to form.  These planets would thus take much longer to accrete than the typical disk lifetime; however, \citet{Kornet:06:661} challenge this result and argue that the probability of planet formation actually \emph{increases} with decreasing stellar mass as a result of a more efficient particle redistribution around lower mass stars.  Another avenue for giant planet formation is the gravitational instability model \citep[e.g.,][]{Boss:97:1836} wherein instabilities in the circumstellar disk can lead to fragmentation and eventual collapse.  \citet{Boss:06:501} argues that gravitational instabilities can produce Jovian worlds around low-mass stars very rapidly.  Current optical RV surveys of M dwarfs support the hypothesis that low-mass stars are less likely to harbor planets than Sun-like stars \citep[e.g.,][]{Endl:06:436, Johnson:10:149}.  Johnson et al.~report that the occurrence rate of giant planets around M dwarfs ($m \sin i > 0.3~M_{JUP}$, $a < 2.5$~AU) is $3.4^{+2.2}_{-0.9}$\% whereas the corresponding fraction for Sun-like stars is 7.6 $\pm$ 1.4\% \citep{Cumming:08:531}.  However, the sample size of planet-hosting M dwarfs is small and therefore the uncertainties are large.  Further exploration of the giant planet population around M dwarfs may therefore offer some insight into which physical processes dominate formation of these worlds.  

Although the investigation of planets around M stars provides clues to their formation, the key to learning about how planets form is to observe them in the process.  Recent results on core accretion predict planet formation in millions of years \citep[e.g.,][]{Alibert:05:343, Hubickyj:05:415, Dodson-Robinson:08:L99}.  Gravitational instability models, however, predict very fast formation times for massive planets in long period orbits: $10^3$ \citep{Mayer:04:1045} to $10^5$ \citep{Bodenheimer:06:1} years.  Subsequent inward migration can also occur on short timescales \citep[$\sim$10$^5$ years,][]{Papaloizou:07:655} that can bring these long period planets into the sensitivity range of RV surveys.  Clearly, any data which can help to identify this timescale will help refine, limit, and possibly eliminate some planet formation models.  Ongoing campaigns to catalog the planets around main sequence stars can not directly provide this information.  By observing late-type, pre-main sequence (PMS) T Tauri stars, one acquires a snapshot of the early stages of planet formation around solar-like stars.  

Low-mass young stars present challenging targets for traditional RV surveys.  Late spectral types, large distances ($>$100 pc), and extinction from natal dust clouds make these targets faint at optical wavelengths.   They also have strong magnetic fields \citep[e.g.,][]{Johns-Krull:07:975} that generate large, cool star spots.   These spots impact RV surveys of young stars by introducing significant jitter which can mimic the RV modulation imposed by a planet \citep{Saar:97:319, Desort:07:983}.  Recent attempts at detecting substellar companions in young stellar populations (10$-$100 Myr) have generally been unsuccessful \citep[e.g.,][]{Paulson:04:3579, Paulson:06:706}, likely because of the small sample size and intrinsic RV variability of the targets.  The youngest RV planet detected to date is a $\sim$6 $M_{JUP}$ planet on an 850 day orbit around the 100 Myr old star HD 70573 \citep{Setiawan:07:L145}.

Spectral line bisector analysis has historically been used to distinguish spot-induced RV variations from companion-induced ones \citep[e.g.,][]{Hatzes:97:374}.  Companion-induced reflex motion does not affect the shape of an absorption line's bisector whereas a spot distorts the line symmetry.  The line bisector span, the difference in bisector values at two different heights of the line profile, is a proxy for the average slope of the line bisector.  A correlation between bisector span and RV variations suggests the RV fluctuation is spot-induced; otherwise, it may be caused by a companion.  However, there may also be no correlation if the projected rotation velocity ($v$ sin $i$) of the star is comparable to or less than the velocity resolution of the spectrograph \citep{Desort:07:983, Prato:08:L103}.  Therefore, bisector analysis is only a first step in identifying potential young planet hosts \citep[e.g.,][]{Huelamo:08:L9, Prato:08:L103}.

A potentially more reliable method for distinguishing between spots and planets leverages the wavelength dependence of the spot-induced RV modulation amplitude.  The reflex motion induced by a planet affects all wavelengths equally.  However, the contrast between a photosphere and a cooler star spot decreases at longer wavelengths because of the flux-temperature scaling in the Rayleigh-Jeans limit of blackbody radiation \citep[e.g.,][]{Vrba:86:199, Carpenter:01:3160}.  As a result, the amplitude of any spot-induced RV variability will be smaller at longer wavelengths.  By observing in the visible \emph{and} NIR it may be possible to distinguish between stellar activity and the presence of a companion by comparing the RV amplitudes at the two wavelengths.  \citet{Reiners:10:432} point out that the magnitude of this decrease for spotted stars is dependent on the temperature difference between the photosphere and star spots and may not be significant if this difference is large (i.e. $\sim$1000 K).  Observations of T Tauri stars, however, do show reductions in RV jitter from optical to K band wavelengths by factors of 3 $-$ 5 \citep{Prato:08:L103}, lending support to the plausibility of this approach.  Furthermore, the late spectral types of T Tauri stars produce SEDs with peak emission around 1 \micron~and extinction of these partly embedded sources is lessened at longer wavelengths, thus allowing for increased signal-to-noise, and hence improved RV precision, in the NIR.   \citet{Reiners:10:432} caution that the advantages of NIR RV measurements may not be apparent for stars earlier than $\sim$M4$-$5.  For early M stars, they assert that the decreased S/N in the optical bands is outweighed by the number of sharp spectral features available at the shorter wavelengths.  Stars later than M5, however, do exhibit increased precision in the NIR both as a result of increased S/N in the J and H bands as well as the appearance of FeH absorption features.  Finally, T Tauri stars allow for some relaxation in the desired RV precision.  While optical surveys of main sequence stars strive for 1 \mps~precision or better, T Tauri stars exhibit an intrinsic NIR RV variability of $>$~100~\mps~\citep{Prato:08:L103, Mahmud:11}.  Therefore, NIR RV precision of several tens of meters per second is more than adequate for these targets.  

Thus, along with studies of planets around M and L dwarfs, RV surveys for young planets also drive the development of improved NIR RV precision measurements.  Several groups are active in this area \citep[e.g.,][]{Martin:06:L75, Blake:07:1198, Huelamo:08:L9, Blake:10:684} with some reports of NIR velocity precision comparable to that in the optical \citep{Bean:10:410, Figueira:10:55}.   While these results are encouraging, these efforts are all focused on large (8 $-$ 10 meter) aperture telescopes.  There is therefore a need to develop precision NIR RV techniques for smaller aperture telescopes where more observing time is available to the community at cadences better tailored to RV surveys.  

In 2004, we began the McDonald Observatory Young Planet Survey to monitor PMS stars in the nearby \citep[$\sim$140 pc,][]{Kenyon:94:1872} Taurus-Auriga low-mass star forming region for evidence of substellar companions.  Our sample of 143 classical and weak-lined T Tauri stars consists of V $<$ 14 stars, most with $v \sin i <$ 20 \kmps \citep{Herbig:88}.  For the first four years, all visible light observations were conducted using the Robert G. Tull Coud{\'e} Spectrometer \citep{Tull:95:251} on the 2.7-meter Harlan J. Smith Telescope.  In 2008, we began to include additional observations from the Sandiford Cassegrain Spectrograph \citep{McCarthy:93:881} on the 2.1-meter Otto Struve Telescope.  Observations at the Kitt Peak National Observatory (KPNO) 4-meter Mayall Telescope using the cassegrain echelle spectrograph were also initiated in late 2008.  Promising targets from these observations, based on measured RV variations and the results of bisector analysis, are selected for follow up K band observations at the 3-meter NASA Infrared Telescope Facility (IRTF) using the high-resolution Cassegrain-mounted echelle spectrograph, CSHELL \citep{Tokunaga:90:131, Greene:93:313}.  Given the 1$-$3 Myr ages of our targets, a positive detection of a young exoplanet will provide a unique datapoint for giant planet formation theory.  

In this paper, we discuss our methodology for measuring precise NIR RVs with CSHELL.  In \S\ref{sec:Observing} we discuss our observing strategy and data reduction algorithms.  In \S\ref{sec:RVs} we present our methodology for using telluric absorption features to measure the RVs of our targets.  We present results from observations of an RV standard star and a known exoplanet in \S\ref{sec:Results}, and we discuss the limitations and possibilities of our technique in \S\ref{sec:Discussion}.

\section{Observation strategy and data reduction}
\label{sec:Observing}
Observations were taken at the 3 meter NASA Infrared Telescope Facility (IRTF) using CSHELL \citep{Tokunaga:90:131, Greene:93:313}.   CSHELL is a long-slit echelle spectrograph (1.08 $\mu$m $-$ 5.5 $\mu$m) that uses a Circular Variable Filter (CVF) to isolate a single order onto a 256$\times$256 InSb detector array.  We used the CVF to isolate a 50 \AA~segment of spectrum centered at 2.298 $\mu$m.  This region contains numerous deep photospheric absorption lines from the CO $\nu$ = 2$-$0 bandhead as well a rich set of predominately CH$_4$ telluric absorption features which we use as a wavelength reference.  The 0\arcsec.5 slit yielded a typical FWHM of 2.6 pixels ($\sim$0.5 \AA, measured from arc lamp spectra) corresponding to a spectral resolving power of R $\sim$ 46,000.   

We observed RV standards, a known exoplanet candidate, and several T Tauri planet host candidates in February 2008 (8 nights), November 2008 (6 nights), February 2009 (2 nights), November 2009 (5 nights), and February 2010 (8 nights).  At the beginning of each night, we imaged twenty flat fields, each with a 20 second integration time, using a continuum lamp to illuminate the entire slit.  We also imaged the same number of 20-second dark frames.  Additionally, we imaged six Ar-Kr-Xe emission lines, changing the CVF while maintaining the grating position, to determine the wavelength reference.  All of our target data were obtained using 10\arcsec~nodded pairs to enable subtraction of sky emission, dark current, and detector bias.  Integration times for each nod were typically 600 seconds; for fainter targets we took multiple contiguous nod pairs.  The signal-to-noise ratio (S/N) for all targets varied significantly depending on cloud cover, seeing, guiding errors, etc., with typical values for all targets of $\sim$50/pixel per nod position.

Our data reduction strategy closely follows that described in \citet{Johns-Krull:99:900} and was implemented entirely in IDL.  We first produced a nightly master dark frame by averaging the twenty individual dark exposures and found a median dark current of 0.24 $e^-$/s.  We produced a nightly normalized flat field by averaging the flat field exposures, subtracting the master dark frame, and then dividing by the mean of the dark-subtracted master flat.  For our target data, we first subtracted the nod pairs to create a difference image.  The difference image was subsequently divided by the normalized flat field.   We estimated the read noise from the standard deviation of a Gaussian fit to a histogram of the pixel values in the difference image ($\sim$30 $e^{-}$).  We then identified the location of the curved spectral traces in the difference image by fitting a second-order polynomial to the location of maximum and minimum flux along the dispersion direction.    

To optimally extract the spectrum, we divided each nod pair into four equally spaced bins of 64 columns along the dispersion direction.  Within each of these 64 column bins we constructed a 10$\times$ oversampled ``slit function" (i.e., the distribution of flux in the cross-dispersion direction).  First, a rough estimate of the spectrum was created by summing the pixels in each column of the difference image for each nod position.  The limits included in this sum are from the midpoint between the two nod positions on the detector to the edge of the area that is well illuminated by the flat lamp. This is typically 60-70 pixels in each column for each nod position. Next, each pixel in the bin was sorted by its distance from the order center for the column the given pixel falls in.  The flux in each pixel was divided by the rough estimate of the spectrum for its appropriate column to normalize all the pixels going into the slit function estimate.  The flux in these offset ordered pixels was then median filtered with a 7 point moving box.  A flux estimate for each oversampled pixel was then determined by taking the median of all the pixels that fell in a given subpixel.  This then formed our oversampled master slit function.  The multiple median filters generally remove the effects of cosmic rays and uncorrected bad pixels on the determination of the slit function.  We then fit this master slit function to a three Gaussian model: a central Gaussian flanked by two satellite Gaussians.  The amplitude, center, and width of each Gaussian were fit as free parameters using the IDL implementation of the AMOEBA non-linear least-squares (NLLS) fitting algorithm \citep{Nelder:65:308}.  The resulting model was then normalized to unit area.  This algorithm produces four model slit functions, one for each bin.  However, the actual slit function is a smoothly varying function of column number.  Therefore, to smooth out the transitions from bin to bin, we create 256 column slit functions by linearly interpolating between the four bin slit functions.  

To determine the total flux in each column of the spectrum, we calculated the scale factor that best matches our model slit function to the column data, per the recipe described in \citet{Horne:86:609}.  In order to mask out spurious flux levels from cosmic rays, we implemented an iterative sigma-clipping algorithm.  This algorithm starts with an estimate of the total noise from the measured read noise in the differenced image plus the Poisson noise from the target.  We then subtracted our model fit from the data in each detector column and masked those pixels whose residual was 3-$\sigma$ greater than the estimated noise.  We iterated through this process 1$-$2 times until the scale factor converged, thus providing an optimal value of the spectrum in that column that is largely immune to hot pixels, cosmic rays, etc.  This algorithm also provides an estimate of the flux uncertainty at each location along the spectrum.

\section{Radial velocity determination}
\label{sec:RVs}
The telluric absorption features in the K-band provide an absolute wavelength and instrumental profile reference, similar in concept to the iodine gas cell technique used in high-precision optical RV exoplanet surveys \citep{Butler:96:500}.  Using the atmosphere as a ``gas cell" lets us superimpose onto our spectra a relatively stable wavelength reference which follows the same optical path as the light from the science target.  This helps alleviate uncertainties introduced by variable slit illumination, changing optical path lengths, etc.  We determined the radial velocities of our targets using a spectral modeling technique similar to the one presented in \citet{Blake:07:1198}, \citet{Prato:08:L103}, and \citet{Figueira:10:55}.  We modeled the stellar spectrum and the telluric features using two high-resolution template spectra.  For the stellar spectrum, we employed NextGen stellar atmosphere models \citep{Hauschildt:99:377} tailored to the $T_{eff}$, $\log g$, and metallicity of our targets.  We used SYNTHMAG \citep{Piskunov:99:515} to generate spectra from the NextGen models along with atomic \citep{Kupka:00:590} and CO \citep{Goorvitch:94:535} line lists.  These spectra are sampled at a resolution $\sim$14 times greater than our observations.  We modeled the telluric features using the NOAO telluric absorption spectrum \citep{Livingston:91} which has a resolution $\sim$4 times higher than that of our observations.  

In order to match the model to our observations, we fit a velocity shift and a power law scaling factor for each template separately.  The power law scaling accounts for differences in line strength between the observed and template spectra \citep[e.g.,][]{McCullough:06:1228}.  The stellar spectrum was then interpolated onto the wavelength scale for the telluric spectrum.  To match stars with measurable rotation, we fit for the value of $v \sin i$ that optimally broadens the stellar template spectrum using a disk-integration routine with a limb-darkening coefficient of 0.2 (based on fitting the model photosphere to a linear limb-darkening law).  For the main sequence stars observed for this paper, however, we fixed $v \sin i$ to zero since it is not meaningfully detected above the instrumental broadening.  Because the normalized telluric spectrum models the relative transmission of the atmosphere, we multiplied the telluric spectrum and broadened stellar spectrum together to create a composite spectrum and convolved the result with a Gaussian instrumental profile (IP) to model the instrument response.  We experimented with multi-Gaussian models of the IP \citep[e.g.,][]{Valenti:95:966, Bean:07:749} but found no improvement in the RV precision.  We then fit the continuum level with a second-order polynomial.  Finally, we binned the 4$\times$ oversampled composite model down to the resolution of our CSHELL data.  The resultant model has nine free parameters (two velocity shifts, two scaling factors, $v \sin i$, IP FWHM, and three continuum coefficients).  To simultaneously determine the best fit to these parameters, we again used the IDL implementation of the AMOEBA NLLS algorithm.

Because CSHELL is cassegrain-mounted, instrument flexure results in a wavelength dispersion on the detector that changes with each observation.  We determined an initial dispersion solution using the Ar-Kr-Xe lamp lines imaged at the start of each night (\S\ref{sec:Observing}).   The six individual arc lamp spectral images were stacked to create a composite spectrum.  The combined arc spectrum was then extracted using the order tracing from that night's highest S/N stellar spectrum.  We found an initial dispersion solution by fitting a third order polynomial to the locations of these emission lines.  This initial polynomial serves as a starting point to refine the dispersion solution.  For each target observation, we take the observed spectrum and the best-fit model and divide both into eight, 6.25 \AA~wide, bins.  This breaks up the observed and model spectra in to eight pieces.  We then take each piece of observed spectrum and cross-correlate it with the corresponding piece of model spectrum to measure the pixel shift between the two.  The pixel shift as a function of spectrum bin provides a crude estimate of how much the observed dispersion solution differs from the initial solution across the detector.  The eight pixel shift values are fit to a second-order polynomial.  By interpolating this polynomial across all detector columns, we estimate by how much the wavelength has shifted per pixel relative to the initial dispersion.  A new wavelength solution is determined by adding this shift to the initial solution.  We then produce a new model fit using the refined wavelength solution.  We iterate this algorithm $\sim$5 times until the wavelength solution converges.  

\begin{figure}
\plotone{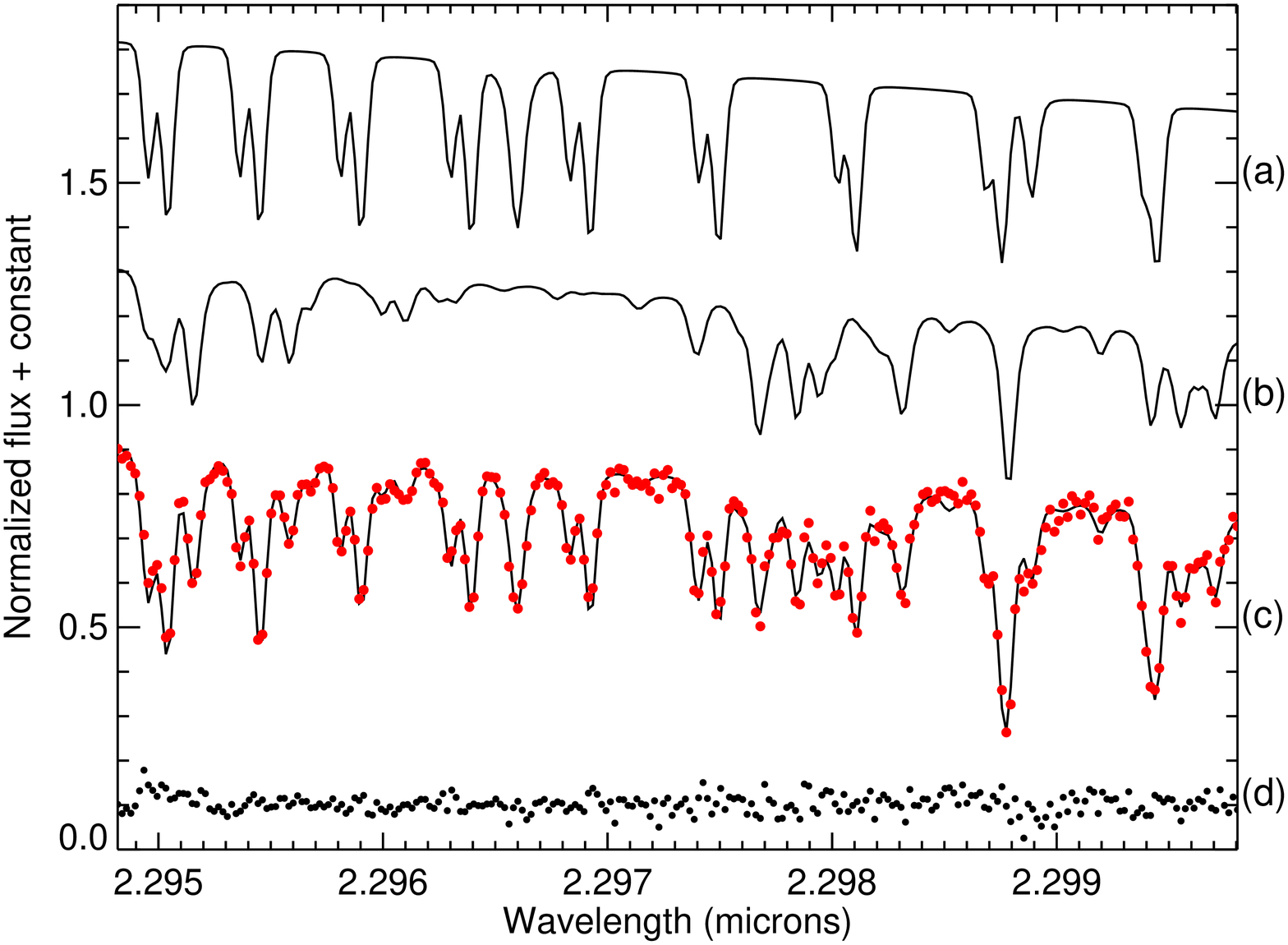}
\caption{Illustration of NIR spectrum modeling for our highest S/N observation of GJ 281 (JD 2455252.906): (a) NextGen photosphere model, (b) telluric template, (c) combined model (solid line) with CSHELL observation (dots), (d) residuals.  All plots are normalized to the final model and have constants added for visual clarity.}
\label{fig:spectrum}
\end{figure}

Figure \ref{fig:spectrum} illustrates the outcome of this modeling for one observation of our RV standard, GJ 281.  Radial velocities for each observation were calculated by taking the difference in the velocity shifts between the best-fit stellar and telluric spectra.  These velocities were in turn corrected for barycentric motion determined for the mid-time of each exposure \footnote{Barycentric corrections were calculated using {\tt helcorr.pro}, an IDL routine based on the IRAF task {\tt noao.astutils.rvcorrect}.}. 

For each observation, we estimated uncertainties introduced by the photon-limited errors of our observations.  These uncertainties in turn have two main components: the first is from the information content in the stellar spectrum itself and the second is in the information content in the telluric spectrum which is used as a wavelength calibrator.  As described in \citet{Butler:96:500}, the intrinsic RV error of a spectrum is the quadrature sum of these two error contributions.  This in turn is a function of the slope of the intensity with respect to wavelength at each pixel and the S/N.  For a given S/N, a spectrum with more numerous and sharper absorption features will have more information content than one with fewer, broader features.  We used the best-fit model photosphere to calculate the intensity derivative and the measured flux uncertainty (estimated from the measured read noise plus the target Poisson noise, \S\ref{sec:Observing}) to determine the S/N for each pixel.  The combined errors from each pixel determine the photon-limited uncertainty on the final RV.  We derived the errors introduced by the telluric spectrum in exactly the same manner, replacing the photosphere model with the telluric one.  The final uncertainty on the velocity was calculated by adding the stellar and telluric errors in quadrature.

A final, nightly RV was determined by calculating the average of the individual nod RVs, weighted by the median S/N across each spectrum.  The uncertainties were computed by taking the weighted standard deviation of the nod RVs and dividing by the square root of the number of nods.  

\section{Results}
\label{sec:Results}
\subsection{GJ 281}
\label{sec:GJ281}
\begin{figure}
\plotone{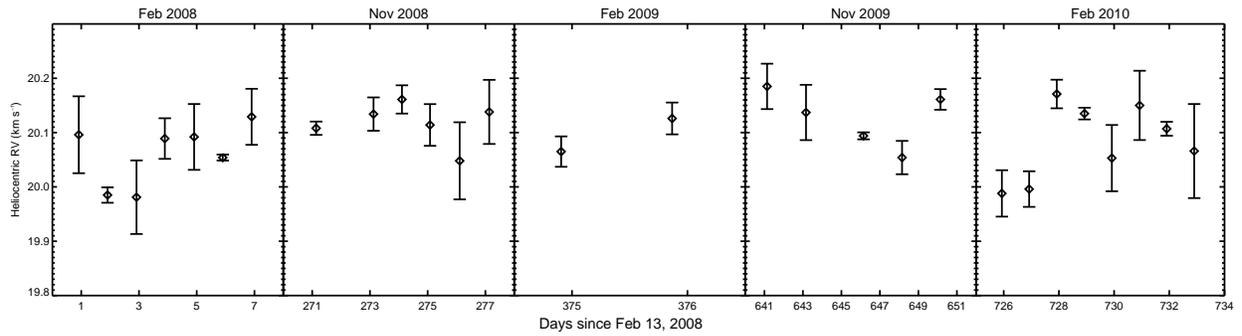}
\caption{Binned K band radial velocities of GJ 281 from Feb 2008 $-$ Feb 2010.  The error bars are the standard deviation of the mean of all nods obtained on that night ($\sigma_{nods}/\sqrt{N}$); they exhibit a median value of $\sim$37 \mps.  The standard deviation of the binned RVs over the entire observing period is 58 \mps.}
\label{fig:gj281}
\end{figure}

To determine the precision of our approach, we obtained 28 observations of GJ 281 over the past two years.  This is a late-type star \citep[M0, K mag = 5.9, $T_{eff}$ = 3776 K,][]{Casagrande:08:585} known to have a stable RV \citep[r.m.s $\sim$ 12 \mps,][]{Endl:03:3099}.  We determined the nightly RVs and uncertainties using the approach described in Section \ref{sec:RVs} (see Figure \ref{fig:gj281} and Table \ref{tbl:gj281}).  The median uncertainty from photon statistics is 40 \mps~per nod with the telluric uncertainty $\sim$50\% greater than the stellar spectrum uncertainty.  The standard deviation of the mean within a sequence of contiguous nod pairs  ($\sigma_{RV}/\sqrt{N}$) is typically 37 \mps.  After averaging, the standard deviation of the nightly RVs over the entire 24 month observing window is $\sim$58 \mps.  Since this is greater than the variability seen in a given night, we need to estimate what additional systematic error we are introducing between nights.  Assuming that the night-to-night systematics add in quadrature with the variability over multiple contiguous nods to give us our overall 58 \mps~standard deviation,  ($\sigma^2_{TOT} = \sigma^2_{nods} + \sigma^2_{nightly}$), we estimate $\sqrt{58^2 - 37^2} = 45$ \mps~as our night-to-night systematic limit.

To better understand the systematic errors, we looked for differences in the measured velocities as a function of nod position.  Subtracting the A beam velocities from the B beam velocities, we found a mean difference between the beams of 66 \mps.  The standard deviation of these differences was 131 \mps, while the standard deviation of the mean was 22 \mps.  The standard deviation of the mean of just the A beam velocities was 15 \mps; for the B beam velocities, this was 17 \mps.  This suggests that there is a statistically significant difference between the velocities at the two nod positions which gets averaged out at some level when combining all observations in a single night.  A simple subtraction from the A velocities to remedy this offset is not sufficient, however.  In a few nod pairs, the B velocities are higher than the A velocities and differencing makes this offset worse.  Because of this, the variability seen in the binned velocities is nearly identical whether or not the mean offset is removed.  

It is not clear why there is a velocity offset between the beams.  While we expect to see different distortions in the two nod positions, the use of telluric features as a common-path calibrator should account for this.  One concern with our observing setup is the asymmetric distribution of telluric lines across the spectrum.  The long wavelength half of our spectra is much richer in telluric features than the short-wavelength half.  It is therefore possible that we are creating false higher order terms as we track the change in the dispersion across the slit.  We tested this by recalculating our RV determination for all the GJ 281 observations using only the left half of the spectrum and then using only the right half of the spectrum.  Fitting only the left half results in a mean difference between the A and B beams of 33 \mps~whereas restricting ourselves to the telluric-rich right half results in a mean difference between the beams of $10^{-6}$ \mps.  While this would seem to support the hypothesis that an asymmetric telluric line distribution may be introducing systematic errors, the standard deviation of the mean in each beam for both cases is large enough to make the statistical significance of these results unimportant.  Interpretation of the much lower RV scatter in the right half of the spectrum is therefore unclear.  It is likely that we simply do not have enough signal to noise to effectively separate out this error from our other sources of uncertainty.  Future observations of very bright targets may help identify the sources of our systematic error.  We note that the existence of a systematic shift between the two beams combined with our use of the standard deviation of the measured RVs in both nod positions will most likely over estimate the uncertainty in our velocity measurements. 

\subsection{Gl 86}
\label{sec:GL86}
\begin{figure}
\plotone{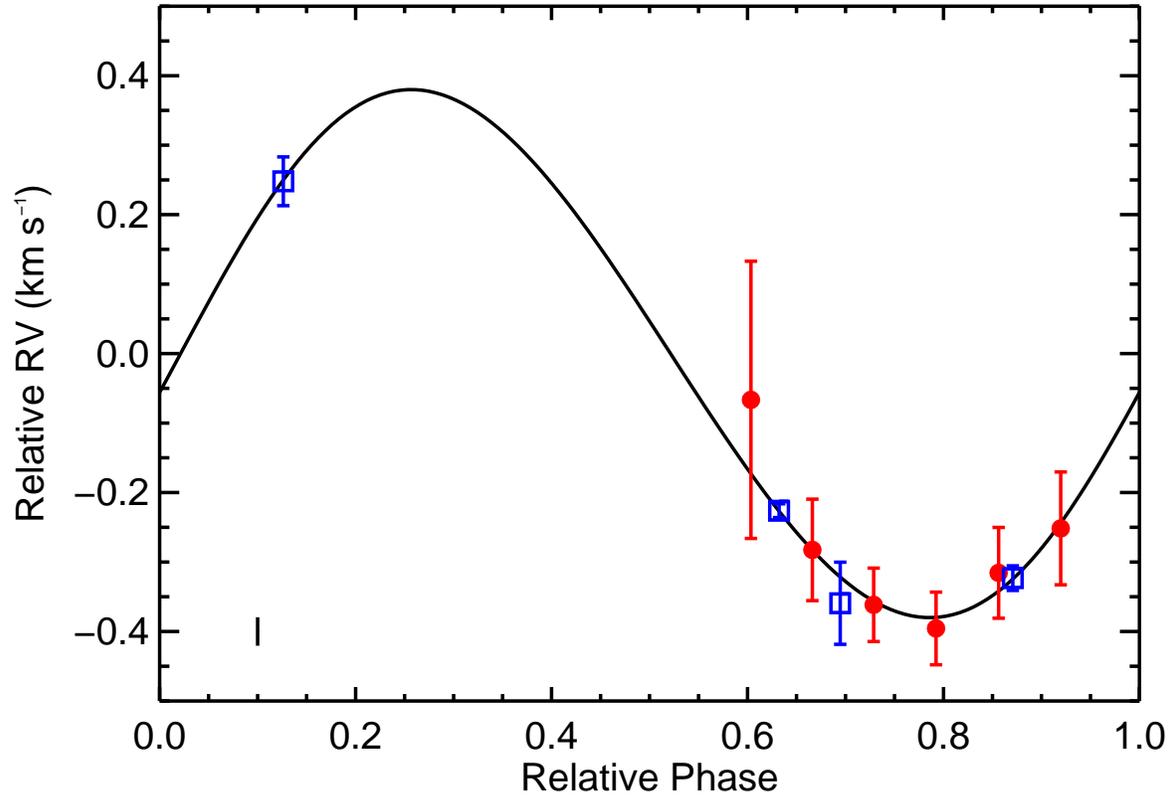}
\caption{Known exoplanet Gl 86 b.  Open blue squares are NIR RVs measured using CRIRES \citep{Figueira:10:55}.  Filled red circles are our binned K band RVs from November 2008.  Black line is orbit model fit to the combined datasets.  Weighted standard deviation of our residuals is 46 \mps.  The bar in the lower left indicates the median uncertainty from photon statistics.}
\label{fig:gl86}
\end{figure}

In addition to RV standards and T Tauri planet host candidates, we also obtained six observations of the known exoplanet host Gl 86 \citep[K mag = 4.1, $T_{eff}$ = 5350 K, $\log g$ = 4.6,][]{Flynn:97:617} in November 2008.  \citet{Queloz:00:99} presented evidence for a hot Jupiter around Gl 86 with $m \sin i$ = 4.0 $M_{JUP}$ and $P = 15.78$ days.  \citet{Figueira:10:55} used Gl 86 as a test case for their precision NIR RV work with the CRIRES spectrograph on the VLT and were able to show that their observations were consistent with the \citeauthor{Queloz:00:99}~orbital solution, allowing for errors in time of periastron ($T_0$) determination and drift in the center-of-mass velocity ($V_r$) as a result of the presence of a long period companion \citep{Eggenberger:03:43}.  We present our nightly RVs for Gl 86 in Table \ref{tbl:gl86} and Figure \ref{fig:gl86}.   The median uncertainty from photon statistics is 41 \mps~with the telluric uncertainty $\sim$43\% greater than the stellar spectrum uncertainty.  The standard deviation of the mean within a sequence of contiguous nod pairs is typically 58 \mps.  Figure \ref{fig:gl86} plots our binned RVs with those from \citeauthor{Figueira:10:55}~and an orbital fit to both datasets.  Our error bars are determined by adding the results from our binning algorithm (\S \ref{sec:RVs}) in quadrature with the 45 \mps~systematic uncertainty determined from our RV standard (\S\ref{sec:GJ281}).  When fitting an orbit model to the data, we allowed $T_0$ and $V_r$ to be fit as free parameters, while fixing all other values to those from \citet{Queloz:00:99}.   We show good agreement with previously published values exhibiting a $\sim$46 \mps~standard deviation in the residuals.  

All of our RVs differ from the orbital solution by no more than 0.5 sigma; the mean difference is $\sim$0.24 sigma.  Assuming normally distributed errors, we would expect 1 $-$ 2 of these residuals to be greater than one sigma.  To explore this further, we assumed that the underlying uncertainties are gaussian and drawn from a distribution with $\sigma = \sqrt{58^2 + 45^2} = 73$ \mps.  The variance in the residuals is dominated by the first RV measurement; excluding that point we find a standard deviation in the residuals of 16 \mps.  We used a Monte Carlo analysis to estimate that the probability of having at least five out of six points within 16 \mps~of the mean, given $\sigma = 73$ \mps, is $\sim$0.08\%. We further explored the false alarm probability (FAP) that the phase coherence from our observations is the result of chance.  We performed a Monte Carlo analysis by holding the times fixed and sampling the CSHELL velocities (and corresponding uncertainties) with replacement over $10^4$ iterations.  For each iteration, we fit an orbit model as described above and recorded how often the reduced $\chi^2$ of the fit was less than that of the fit to the observed velocities.  We find that it is highly unlikely that the phase coherence is a result of chance, with a FAP of 0.0003.  We are therefore very confident that our RVs are consistent with the known planetary-mass object orbiting Gl 86.  The inconsistency between our estimated error bars and the residual scatter is suspicious, however.  The scatter in the residuals is more comparable to the estimated photon uncertainty.  Our 45~\mps~systematic uncertainty is measured over a period of two years whereas the RVs for Gl 86 were measured over a single run.  If the systematics were ``quiet" on this run, then 45~\mps~may not be the appropriate value to use.  We see some evidence for the appearance of ``quiet" times in our data on GJ 281.  During the November 2008 run the standard deviation of our nightly (6 nights) GJ 281 data is 39 \mps, well below the 58 \mps~of the entire data set.  This is the same run on which the Gl 86 data were obtained.  We therefore believe that our uncertainties are most likely overestimates.  

\section{Discussion}
\label{sec:Discussion}
NIR spectroscopy has only recently begun to play a role in the search for substellar companions.  \citet{Martin:06:L75} combined optical and H band observations to look for giant planets around the young brown dwarf LP 944-20; they concluded that the observed optical RV modulations were driven by inhomogeneous surface features (i.e. clouds).  \citet{Blake:07:1198} used high resolution K-band spectroscopy to investigate the presence of giant planets around a population of L dwarfs, achieving a precision of 300 \mps.  They find no evidence for companions with $M \sin i > 2 M_J$ and $P < 3$ days in their sample of nine targets.  \citet{Setiawan:08:38} reported the detection of a $\sim$10 $M_J$ planet on a 3.5 day orbit around the 10 Myr old star TW Hydra. However, H band observations presented by \citet{Huelamo:08:L9} reveal a strong wavelength dependence in the velocity amplitude, thus casting doubt on the presence of a companion and suggesting that spots are the cause of TW Hydra's RV variations.  \citet{Prato:08:L103} observed two potential planet-host candidates in the NIR, selected on the basis of their optical variability.  To within their measurement precision, they detect no K band RV modulation, implying that spots are responsible for the apparent optical RV variability.  

The aforementioned interest in targeting M dwarfs as hosts of habitable planets (\S\ref{sec:Intro}) has spurred vast improvements in NIR precision, bringing it closer to that of optical surveys.  \citet{Bean:10:410} report achieving a long term precision in the H band of $\sim$5 \mps~on late M dwarfs using CRIRES at the VLT with the aid of an ammonia gas cell.  \citet{Figueira:10:55} report a comparable precision also on CRIRES but using telluric absorption features as a wavelength reference. Based on six years of NIRSPEC data, \citet{Blake:10:684} report a precision of 50 \mps~for their sample of K dwarfs and 200 \mps~for L dwarfs also through the use of telluric lines.  

In an earlier paper from our survey of young, low-mass stars, \citet{Prato:08:L103} presented evidence that the RV variability of two T Tauri stars, DN Tau and V836 Tau, is the result of star spots.  These two targets exhibited an optical RV standard deviation of 438 \mps~and 742 \mps, respectively.  K band RVs, however, showed a standard deviation of 144 \mps~for DN Tau and 149 \mps~for V836 Tau.  These values are consistent with the measurement uncertainties of $\sim$165 \mps~and $\sim$290 \mps, respectively.  These stars therefore exhibit no evidence of RV jitter in the K band.  The undetected NIR RV variability demonstrates two important points.  First, optical RV surveys can not rely on bisector analysis alone to ascertain the mechanism for RV variability.  Both DN Tau and V836 Tau  showed significant periodic optical variability and neither exhibited a correlation between RV and bisector span.  \emph{Multiwavelength observations are therefore an essential element for testing the planet hypothesis}, especially in young stars.  Second, they support the hypothesis that RV jitter from stellar activity should be significantly lower at longer wavelengths.  We have demonstrated $\sim$58 \mps~precision with CSHELL over two years and the ability to detect a known 4 $M_{JUP}$ planet.  If the $\sim$150 \mps~variability observed in DN Tau and V836 Tau is typical of T Tauris, our 58 \mps~precision is well below the intrinsic noise floor of our targets suggesting that our methodology is well-suited to detecting ``hot Jupiters" around young stars.  However, we caution against extrapolating from two young stars to the NIR behavior of other T Tauri stars.  \citet{Mahmud:11} present another target from our Taurus survey that exhibits a $\sim$430 \mps~modulation in the K band, which we believe to be caused by spots, suggesting that the NIR RV behavior of T Tauris may vary significantly from star to star.

\acknowledgments
The authors thank R. White and J. Bailey for productive discussions on data reduction and our anonymous referee for offering many useful suggestions which improved the manuscript.  We acknowledge the \emph{SIM} Young Planets Key Project for research support; funding was also provided by NASA Origins Grants 05-SSO05-86 and 07-SSO07-86.  This work made use of the SIMBAD database, the NASA Astrophysics Data System, and the Two Micron All Sky Survey (2MASS), a joint project of the University of Massachusetts and IPAC/Caltech, funded by NASA and the NSF.  We recognize the significant cultural role that Mauna Kea plays in the indigenous Hawaiian community and are grateful for the opportunity to observe there.

{\it Facilities:} \facility{IRTF (CSHELL)}.

\bibliography{articles,notarticles}

\clearpage

\clearpage

\clearpage

\clearpage
\begin{deluxetable}{ccc}
\tablewidth{0pt}
\tablecaption{GJ 281 Radial Velocities}
\tablehead{
\colhead{JD - 2450000} & \colhead{Heliocentric RV (\kmps)} & \colhead{$\sigma_{nods}/\sqrt{N}$ (\mps)}
}
\startdata
4510.914 & 20.096 &  70.8 \\
4511.910 & 19.985 &  14.2 \\
4512.914 & 19.981 &  67.7 \\
4513.895 & 20.089 &  37.4 \\
4514.910 & 20.092 &  60.6 \\
4515.902 & 20.054 &   5.5 \\
4516.902 & 20.129 &  51.6 \\
4781.137 & 20.108 &  12.3 \\
4783.125 & 20.134 &  30.7 \\
4784.109 & 20.161 &  26.0 \\
4785.082 & 20.114 &  38.4 \\
4786.109 & 20.048 &  71.0 \\
4787.133 & 20.138 &  59.0 \\
4884.906 & 20.065 &  28.0 \\
4885.867 & 20.126 &  29.3 \\
5151.137 & 20.185 &  41.8 \\
5153.160 & 20.137 &  50.9 \\
5156.148 & 20.094 &   6.4 \\
5158.152 & 20.054 &  30.7 \\
5160.145 & 20.161 &  19.0 \\
5235.934 & 19.988 &  42.7 \\
5236.922 & 19.996 &  32.8 \\
5237.922 & 20.171 &  26.3 \\
5238.926 & 20.135 &  10.8 \\
5239.918 & 20.053 &  61.0 \\
5240.926 & 20.150 &  63.8 \\
5241.906 & 20.107 &  12.8 \\
5242.902 & 20.066 &  86.6 \\
\enddata
\label{tbl:gj281}
\end{deluxetable}

\clearpage
\begin{deluxetable}{ccc}
\tablewidth{0pt}
\tablecaption{Gl 86 Radial Velocities}
\tablehead{
\colhead{JD - 2450000} & \colhead{Heliocentric RV (\kmps)} & \colhead{$\sigma_{nods}/\sqrt{N}$ (\mps)}
}
\startdata
4781.902 & 56.303 & 199.5 \\
4782.891 & 56.087 &  73.0 \\
4783.879 & 56.008 &  52.8 \\
4784.883 & 55.974 &  52.3 \\
4785.891 & 56.054 &  65.4 \\
4786.891 & 56.118 &  81.2 \\
\enddata
\label{tbl:gl86}
\end{deluxetable}

\end{document}